\documentstyle[11pt,epsf]{article}

\newcommand{\beq}{\begin{equation}}
\newcommand{\eeq}{\end{equation}}
\newcommand{\beqa}{\begin{eqnarray}}
\newcommand{\eeqa}{\end{eqnarray}}

\newcommand{\NPB}[1]{{\it Nucl. Phys.}\ {\bf B{#1}}}
\newcommand{\PLB}[1]{{\it Phys. Lett.}\ {\bf B{#1}}}
\newcommand{\PRD}[1]{{\it Phys. Rev.}\ {\bf D{#1}}}
\newcommand{\PRL}[1]{{\it Phys. Rev. Lett.}\ {\bf #1}}

\begin{document}

\begin{titlepage}
\def\thepage {}        

\title{Custodial Symmetry and the Triviality Bound \\
on the Higgs Mass}

\author{
R.Sekhar Chivukula\thanks{e-mail addresses: sekhar@bu.edu, simmons@bu.edu},
and Elizabeth H. Simmons$^*$\\
Department of Physics, Boston University, \\
590 Commonwealth Ave., Boston MA  02215}

\date{August 1996}

\maketitle

\bigskip
\begin{picture}(0,0)(0,0)
\put(295,250){BUHEP-96-26}
\put(295,235){hep-ph/9608320}
\end{picture}
\vspace{24pt}

\begin{abstract}

  The triviality of the scalar sector of the standard one-doublet Higgs
  model implies that it is only an effective low-energy theory valid
  below some cut-off scale $\Lambda$.  In this note we show that the
  experimental constraint on the amount of custodial symmetry violation,
  $\vert\Delta \rho_* \vert = \alpha \vert T \vert \le 0.4\%$, implies
  that the scale $\Lambda$ must be greater than of order 7.5 TeV. For
  theories defined about the infrared-stable Gaussian fixed-point, we
  estimate that this lower bound on $\Lambda$ yields an upper bound of
  approximately 550 GeV on the Higgs boson's mass, independent of the
  regulator chosen to define the theory. We also show that some
  regulator schemes, such as higher-derivative regulators, used to
  define the theory about a different fixed-point are particularly
  dangerous because an infinite number of custodial-isospin-violating
  operators become relevant.

\pagestyle{empty}
\end{abstract}
\end{titlepage}


\section{Introduction}

The triviality \cite{triviality} of the scalar sector of the standard
one-doublet Higgs model implies that this theory is only an effective
low-energy theory valid below some cut-off scale $\Lambda$.
Physically this scale marks the appearance of new strongly-interacting
symmetry-breaking dynamics. As the Higgs mass increases, the upper
bound on the scale $\Lambda$ decreases.  If we require that $M_H /
\Lambda$ be small enough to afford the effective Higgs theory some
range of validity (or to minimize the effects of regularization in the
context of a calculation in the scalar theory), one arrives at an
upper bound on the Higgs boson's mass
\cite{cabbibo, dashen}.

Quantitative studies on the lattice using analytic \cite{luscher} and
Monte Carlo \cite{kuti, hasenfratz, bhanot, gockeler, heller} techniques
result in an upper bound of approximately 700 GeV.  However, these
lattice results are potentially ambiguous because the precise value of
the bound on the Higgs boson's mass depends on the restriction placed on $M_H /
\Lambda$. The ``cut-off'' effects arising from the regulator are not
universal: different schemes can give rise to different effects of
varying sizes and can change the resulting Higgs mass bound.

In this note we show that, for models that reproduce the standard
one-doublet Higgs model at low energies, electroweak phenomenology
provides a lower bound on the scale $\Lambda$ that is
regularization-independent (i.e. independent of the details of the
underlying physics). Recall that the standard one-doublet Higgs model
has an accidental custodial isospin symmetry
\cite{custodial}, which naturally implies that the weak-interaction
$\rho$-parameter is approximately one. While all $SU(2)
\times U(1)$ invariant operators made of the scalar-doublet field that have
dimension less than or equal to four automatically respect custodial
symmetry, terms of higher dimension that arise from the underlying
physics at scale $\Lambda$ in general will not. We show that current
results from precision electroweak tests \cite{something}, which
provide the constraint 
\beq
\vert \Delta \rho_*\vert \le 0.4\%
\eeq
on $\Delta \rho_*$ ($= \alpha T$) \cite{rho,pt} at the 95\% confidence level,
imply that the scale $\Lambda$ must be greater than approximately 7.5
TeV. This lower bound on $\Lambda$ implies that the Higgs boson's mass
must be less than approximately 550 GeV, independent of the cut-off
method chosen to define the theory.

Implicitly assumed in these bounds is the naive scaling that one
expects near the infrared-stable Gaussian fixed point of scalar field
theory. Other fixed points with very different scaling behavior may
also exist.  Typically, these new fixed points correspond to
field theories with an infinite number of relevant operators
\cite{kutii}.  A nice example of this possibility has been
explored by Jansen, Kuti, and Liu \cite{jansen}, who performed an
analytic (large-N) analysis of the Higgs model in the presence of a
pair of complex-conjugate Pauli-Villars regulator fields.  Their
calculations show the possibility of defining the theory with a Higgs
mass of 2 TeV (!) while forcing the ghost (Pauli-Villars) states to
have masses greater than 4 TeV. However, we will show that in this
theory there are an infinite number of relevant
custodial-isospin-violating operators.  Therefore, given the degree of
custodial isospin violation present in the splitting between the
masses of the top and bottom quarks, these theories cannot give rise
to phenomenologically viable theories of a heavy Higgs boson. We
expect that our results will generalize to other potentially
``non-trivial'' scalar field theories as well.

\section{Triviality and custodial symmetry}

We begin by considering an underlying theory which is arbitrary and
does not respect custodial symmetry.  We are interested in cases which
reproduce the standard one-doublet Higgs model at low energies.  The
low-energy theory should respect $SU(2)_W \times U(1)_Y$ and the only
low-energy state resulting from the underlying dynamics should
be the Higgs doublet.  Since we are considering theories with a heavy
Higgs field, we expect that the underlying high-energy theory will be
strongly interacting.

To estimate the sizes of various effects of the underlying physics, we
will rely on dimensional analysis. As noted by Georgi
\cite{generalized}, a theory\footnote{These dimensional estimates only
apply if the low-energy theory, when viewed as a scalar field theory,
is defined about the infrared-stable Gaussian fixed-point. We return
to potentially ``non-trivial'' theories below.}  with light scalar
particles belonging to a single symmetry-group representation depends
on two parameters: $\Lambda$, the scale of the underlying physics, and
$f$ (the analog of $f_\pi$ in QCD), which measures the amplitude for
producing the scalar particles from the vacuum. Our estimates will
depend on the ratio $\kappa = \Lambda / f$, which is expected to fall
between 1 and $4\pi$. For example, in a QCD-like theory with $N_c$
colors and $N_f$ flavors one expects
\cite{reconsider} that
\beq
\kappa \approx \min \left({4\pi a\over N_c^{1/2}},
{4\pi b\over N_f^{1/2}}\right)~,
\eeq
where $a$ and $b$ are constants of order 1. In the results that
follow, we will display the dependence on $\kappa$ explicitly; when
giving numerical examples, we set $\kappa$ equal to the geometric mean
of 1 and $4\pi$, {\it i.e.} $\kappa \approx 3.5$.

Because of the $SU(2)_W \times U(1)_Y$ symmetry of the low-energy
theory, all terms of dimension less than or equal to four respect
custodial symmetry \cite{custodial}. The leading custodial-symmetry
violating operator is of dimension six \cite{wyler,
grinstein} and involves four Higgs doublet fields
$\phi$. According to the rules of dimensional analysis, the
operator
\beq
{\lower35pt\hbox{\epsfysize=1.0 truein \epsfbox{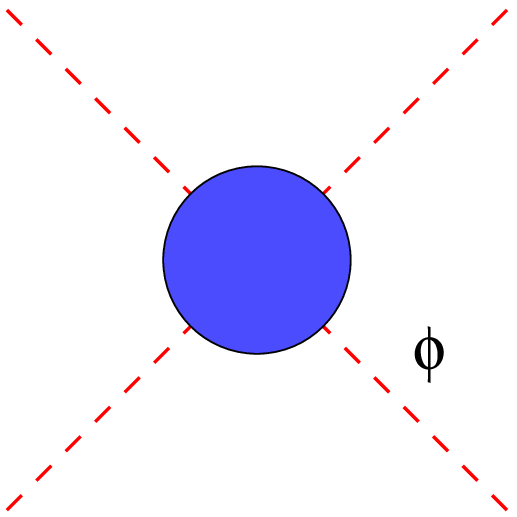}}}
\Rightarrow
{\kappa^2 \over \Lambda^2} 
(\phi^\dagger D^\mu \phi)
(\phi^\dagger D_\mu \phi)~,
\label{eq:isoviol}
\eeq
should appear in the low-energy effective theory with a coefficient
of order one \cite{grinstein}. Such an operator will give rise to a deviation
\beq
\Delta \rho_* = - {\cal O}\left(\kappa^2 {v^2 \over \Lambda^2}\right) ~,
\eeq
where $v \approx 246$ GeV is the expectation value of the Higgs
field. Imposing the constraint \cite{something} 
that $|\Delta \rho_*| \le 0.4\%$, we find the lower bound
\beq
\Lambda \stackrel{>}{\sim} 4\, {\rm TeV} \cdot \kappa ~.
\eeq
For $\kappa \approx 3.5$, we find $\Lambda \stackrel{>}{\sim}
14$ TeV. 

Alternatively, it is possible that the underlying strongly-interacting
dynamics respects custodial symmetry. Even in this case, however,
there must be custodial-isospin-violating physics (analogous to
extended-technicolor interactions \cite{etc}) which couples the $\psi_L=
(t,\ b)_L$ doublet and $t_R$ to the strongly-interacting ``preon''
constituents of the Higgs doublet in order to produce a top quark
Yukawa coupling at low energies and generate the top quark mass. If,
for simplicity, we assume that these new weakly-coupled
custodial-isospin-violating 
interactions are gauge interactions with coupling $g$ and mass $M$,
dimensional analysis allows us to estimate the size of the resulting
top quark Yukawa coupling
\beq
{\lower35pt\hbox{\epsfysize=1.0 truein \epsfbox{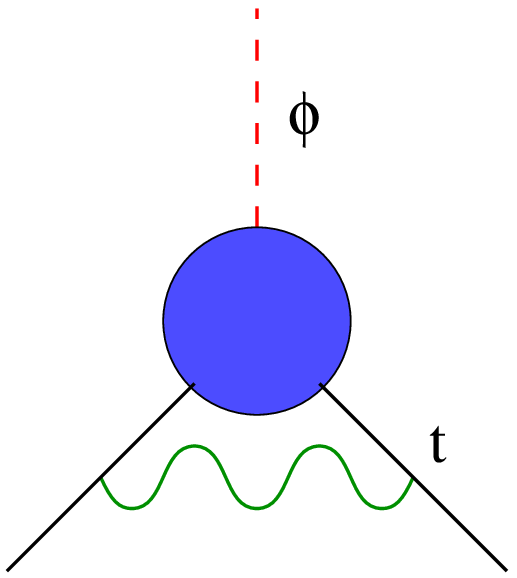}}}
\Rightarrow
{g^2 \over M^2} {\Lambda^2 \over \kappa}\bar{t}_R \phi \psi_L ~.
\eeq
In order to give rise to a suitably large top-quark mass 
the Yukawa coupling must be greater than or of order one, implying that
\beq
\Lambda \stackrel{>}{\sim} {M \over g} \sqrt{\kappa}~.
\label{eq:yukawa}
\eeq 
These new gauge interactions will typically also give rise to
custodial-isospin-violating 4-preon interactions\footnote{These
  interactions have previously been considered in the context of
  technicolor theories\cite{appelquist}.} which, at low energies, will
give rise to an operator of the same form as the one in eqn.
\ref{eq:isoviol}. Using dimensional analysis, we find
\beq
{\lower35pt\hbox{\epsfysize=1.0 truein \epsfbox{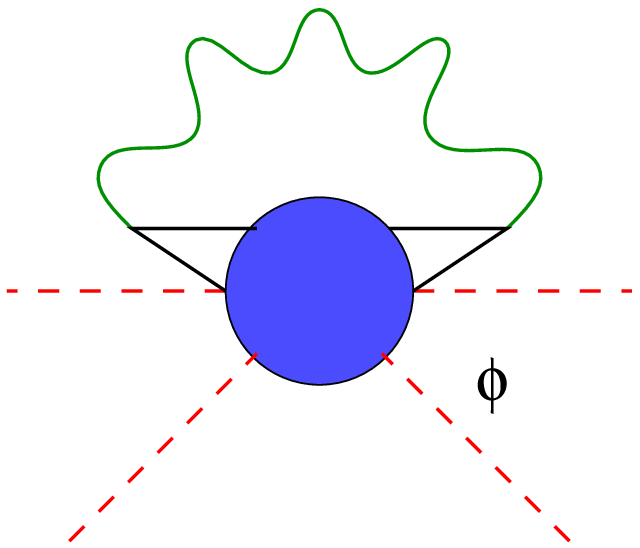}}}
\Rightarrow {g^2 \over M^2} (\phi^\dagger D^\mu \phi) (\phi^\dagger
D_\mu \phi)~,
\label{eq:isoviola}
\eeq
which results in the bound  $M/g \stackrel{>}{\sim} 4$ TeV.
From eqn. \ref{eq:yukawa} we then derive the limit
\beq
\Lambda \stackrel{>}{\sim} 4\, {\rm TeV} \cdot \sqrt{\kappa}~.
\eeq
For $\kappa \approx 3.5$, we find $\Lambda \stackrel{>}{\sim}
7.5$ TeV. 

Because of triviality, a lower bound on the scale $\Lambda$ yields an
upper-bound on the Higgs boson mass. A rigorous result would require a
nonperturbative calculation of the Higgs mass in an $O(4)$-symmetric
theory subject to the constraint that $\Lambda / v \stackrel{>}{\sim}
30$. Here we provide an estimate of this upper bound by naive
extrapolation of the lowest-order perturbative result\footnote{Though
not justified, the naive perturbative bound has been remarkably close
to the non-perturbative estimates derived from lattice Monte Carlo
calculations \cite{kuti, hasenfratz, bhanot, gockeler, heller}.}.
Integrating the lowest-order beta function for the Higgs self-coupling
$\lambda$
\beq
\beta(\lambda) = \mu{d\lambda\over d\mu} = {3\over 2\pi^2}\lambda^2
+\ldots~,
\eeq
we find
\beq
{1\over \lambda(\mu)} - {1\over \lambda(\Lambda)} 
= {3\over 2\pi^2} \log{\Lambda\over \mu} ~.
\eeq
Using the relation $m^2_H = 2\lambda(m_H) v^2$
we find the relation
\beq
m^2_H \log\left({\Lambda\over m_H}\right)\le {4\pi^2 v^2 \over 3}~.
\eeq
For $\Lambda \stackrel{>}{\sim} 7.5$ TeV, this results in the
bound\footnote{If $\kappa \approx 4\pi$, $\Lambda$ would have to be
  greater than 14 TeV, yielding an upper bound on the Higgs boson's mass
  of 490 GeV. If $\kappa \approx 1$, $\Lambda$ would be greater than 4
  TeV, yielding the upper bound $m_H \stackrel{<}{\sim} 670$ GeV.}
$m_H \stackrel{<}{\sim} 550$ GeV.

\section{Non-trivial scaling behavior}

Dimensional analysis was crucial to the discussion given above.  If the
low-energy Higgs theory does not flow toward the trivial Gaussian
fixed-point in the infrared limit, the scaling dimensions of the fields
and operators can be very different than naively expected.  In this case
the bounds given above do not apply.

A nice example of a scalar theory with non-trivial behavior has been
given by Jansen, Kuti, and Liu \cite{jansen}. They
consider a theory defined by an  $O(4)$-symmetric Lagrange-density
with a modified kinetic-energy
\beq
{\cal L}_{kin} = -{1\over 2} \phi^\dagger (\Box + {\Box^3\over {\cal M}^4})
\phi~.
\eeq
In the large-$N$ limit, this higher-derivative kinetic term is
sufficient to eliminate all divergences. A lattice simulation of this
theory \cite{liu} indicates that this approach can be used to define a
non-trivial Higgs theory with a Higgs boson mass as high as 2 TeV, while
avoiding any noticeable effects from the (complex-conjugate) pair of
ghosts which are present because of the higher derivative kinetic-energy
term. 

As shown by Kuti \cite{kutii}, in the infrared this higher-derivative
theory flows to a non-trivial fixed point on an infinite dimensional
critical surface, which corresponds to a continuum field theory with an
infinite number of relevant operators. The reason there are an infinite
number of relevant operators is that, if the continuum limit is taken so
that the scale ${\cal M}$ remains finite as required in order to flow
to a non-trivial theory, the scaling dimension \cite{kutii} of the
Higgs doublet field $\phi$ is -1 instead of the canonical value of
+1! 

If one could impose an exact $O(4)$ symmetry on the symmetry breaking
sector, this would lead to a strongly-interacting electroweak
symmetry-breaking sector without technicolor \cite{liu}.  However, as
argued above, custodial isospin violation in the flavor sector must
couple to the 
symmetry-breaking sector to give rise to the different top- and
bottom-quark masses. Furthermore, if the scaling dimension of the
Higgs field is -1, there is an infinite class of custodial-isospin-violating
operators (including the operator in eqn. \ref{eq:isoviol}) which are
relevant. Since these operators are relevant, even a small amount of 
custodial isospin violation coming from high-energy flavor dynamics will
be amplified as one scales to low energies, ultimately contradicting
the bound on $\Delta\rho_*$. We therefore conclude
that these non-trivial scalar theories cannot provide a phenomenologically 
viable theory of electroweak symmetry breaking. 

To construct a phenomenologically viable theory of a
strongly-interacting Higgs sector it is not sufficient to simply
construct a theory with a heavy Higgs boson, one must also ensure that
all potentially custodial-isospin-violating operators 
remain irrelevant\footnote{This is also a concern in walking technicolor
\cite{rsc}.}.

\section{Conclusions}

We have shown that theories with a heavy Higgs boson which reproduce the
standard model at low energies are caught between the rock of the $\rho$
parameter and the hard place of the top-bottom mass splitting.  In
theories which flow to the infrared-stable Gaussian fixed point, the
scale of the new strongly-interacting dynamics must be greater than of
order 7.5 TeV, and therefore the Higgs boson must weigh less than
approximately 550 GeV.  This result applies whether the
strongly-interacting preon dynamics that underlies the Higgs state
conserves or violates custodial isospin.  In theories with non-trivial
scaling behavior, the presence of an infinite class of relevant
custodial-isospin-violating operators makes it impossible to both
provide a top quark mass and obey the bound on $\Delta \rho_*$.  Such
theories cannot, therefore, provide a phenomenologically acceptable
description of electroweak symmetry breaking.

\bigskip
{\bf Authors' Note}  After the completion of this work, it was brought
to our attention that the possibility of using the $\rho$ parameter as
an additional handle on limiting the Higgs self-coupling had been
suggested, but not pursued, in \cite{neub}.


\vspace{12pt} \centerline{\bf Acknowledgments} \vspace{12pt}

We thank the Fermilab Summer Visitors Program for hospitality (and the
Fermilab Theory Group for a memorable volleyball game) during the
initiation of this work.  We also thank John Terning and Bogdan Dobrescu
for comments on the manuscript. R.S.C. acknowledges the support of the
NSF Presidential Young Investigator program. E.H.S. acknowledges the
support of the NSF Faculty Early Career Development (CAREER) program and
the DOE Outstanding Junior Investigator program. {\em This work was
  supported in part by the National Science Foundation under grants
  PHY-9057173 and PHY-9501249, and by the Department of Energy under
  grant DE-FG02-91ER40676.}


\end{document}